\documentclass[allclo,superscriptaddress,eqsecnum,amsfonts,showpacs]{revtex4}
\usepackage{epsfig}

\newcommand{\be}{\begin{equation}}
\newcommand{\ee}{\end{equation}}
\newcommand{\bea}{\begin{eqnarray}}
\newcommand{\eea}{\end{eqnarray}}

\newcommand{\ep}{i\varepsilon}
\newcommand{\nn}{\nonumber}

\begin{document}

\preprint{ \parbox{1.5in}{\leftline{hep-th/??????}}}

\title{Running Top quark mass in the presence of light SM Higgs}

\author{V.~\v{S}auli}
\affiliation{CFTP and Departamento de F\'{\i}sica,
Instituto Superior T\'ecnico, Av. Rovisco Pais, 1049-001 Lisbon,
Portugal }

\begin{abstract}
The running of the Top quark mass is considered in the nonperturbative framework of the Schwinger-Dyson equation. Based on the input of  physical pole mass meassured at the Tevatron the method provides the resulting mass function which is almost constant at low spacelike and timelike scales.
The skeleton loops including Standard Model Higgs and gluons are taken into account.
The dominant two-loop skeleton contribution with triplet Higgs interaction is considered in addition to one loop dressed approximation of the top quark self-energy.   
\end{abstract}

\pacs{11.10.St, 11.15.Tk}
\maketitle
%

%%%%%%%%%%%%%%%%%%%%%%%%%%%%%%%%%%%%%%%%%%%%%%%%%%%%%%%%%%%%%%%%%%%%%%%%%%%%%%%%%%%%%

\section{Introduction}

The quark masses are fundamental parameters of the Standard Model. The precise knowledge of
quark masses at various scales is important for several reasons. In hadronic physics
it is necessary  for precise determination of CKM matrix elements, while the theoretically extracted information about values of quark masses at very high momenta  can be useful for model builders.

In perturbation theory QCD approach the definition of the running mass is based on the renormgroup evolution equations (RGEs).  MS bar scheme represents short range mass definition and is commonly used  due to its technical simplicity \cite{FUSKOI1997,CHESTE2000}. On the other side the RGEs method cannot provide reliable results at low momenta  where the perturbation method fails.
A legitimate question is what is the relevant scale of the applicability of the perturbation theory, when the corrections to the quark masses are evaluated.
To determine this, the running masses calculation has to rely on nonperturbative QCD techniques. So far, there are two methods that directly follow from the first principles:  the first is  lattice theory (for a review see \cite{LUB1999,RYAN2001,LEL2003,SHOONO2004,RAKOW2004,lat2007}) which  is based on the discretized Euclidean space, the second is the functional method represented  by  a continuous framework of QCD Schwinger-Dyson equations \cite{ALKSME2001,ROBERTS}. Within  certain phenomenological assumptions the QCD sum rules \cite{SUM1,SUM2} are used to determine the quark masses.

After the top quark discovery, the top quark mass value is obtained in the fairly limited regime, the CDF \cite{CDF}
and DO \cite{D0} collaborations measure the resonant top quark mass. The particle data group \cite{PDG} quoted the value $M_{t}=174.2\pm3.3 GeV$ as the pole mass of the top quark. 

The evolution of the Yukawa coupling has already been studied  in \cite{FUSKOI1997}. However, the Yukawa interaction has not been considered in RGEs for the top quark running mass.  The contribution to the quark selfenergy due to the Higgs boson has been studied in  SDE framework for the first time in \cite{SMJAKA1994}.  This  study has been performed with two approximately equivalent inputs:  $M(0)=179GeV$ and $M(M_t^2)=174GeV$, noting that the later was defined incorrectly at  spacelike scale $M_t$. In the present paper we go beyond the one loop approximation and take into account the two loop Higgs contribution as well.

There is a striking evidence that the RGE perturbation calculation overestimated low scale top quark mass from the very beginning (going from high spacelike $Q^2$ to the infrared values). Recall that in the paper \cite{FUSKOI1997} the renormgroup equation for top quark running mass $M(\mu^2)$  has been solved in MS bar renormalization scheme with the following result (in GeV):  
\bea
&&M(180^2)=170.1; \, M(M^2)=170.8;  \, M(91.2^2)=180; \,
\nn \\ 
&&M(4.3^2)=253; \, M(1.3^2)=318; \, M(1)=339 \, ,
\eea
for spacelike arguments in the brackets and  have  been quoted within $\simeq (12-25) GeV$ error due to  experimentally  determined physical mass ( to that date, it was $M_{t}=180GeV$).

Recall that the physical pole mass $M_t $ is determined in Minkowski space as
the $S^{-1}(M_{t})=0$, in other words  $M_{t}(-M_{t}^2)=M_{t}$.  Assuming that   
the fit procedure of $M(M_{t}^2)$ from $M_{t}$ used in \cite{FUSKOI1997}   and developed originally in  \cite{GRAY} is reliable (note, the relation between MS mass and on shell mass is recently known to the order $\alpha^3$), one necessarily must conclude that the renormgroup evaluation of masses becomes unreliably overestimated below the scale $\mu\simeq M_z$. While for leptons and light quarks  perturbation QCD    works perfectly at the $M_z$ scale,
it appears  that for an accurate estimate of  the running top quark mass at
$M_z$ mass scale might not be adequate. Technically this is because already  one loop correction to $M$ is enhanced like 
\be 
\delta M \sim \alpha_{QCD}M \, .
\ee
In other words,  the exceptionally large mass of the top quark itself  
spoils the usual correctness of perturbative QCD at electroweak scale.

This is one of the main reason of the present study to calculate the evolution of top quark mass in the whole  momentum of range, obtaining thus correct information for the low energy scales.
In perturbative MS schemes the running mass grows from MS  value $m(m)=170 $ to $\mu=M_Z$ 
about amount of $10 GEV$ and further blows up when evolved to the infrared. We will argue that $M(M_t)$ and $M(M_Z)$ differs about tiny  amount $\simeq 1-2GeV $ and the running top quark mass function remains stable when using selfconsistent framework of our SDE equations.
The knowledge of here observed infrared stability of the top quark mass should be useful whenever a selfconsistent treatment  is  required, i.e. for instance when one considers Higgsonia \cite{GRITRO2007,KONF} and the effect of top quark loop in the equations for Higgsonium bound states. 
Further, the top quark circulates in the loop of penguin diagrams describing  rare mesonic electroweak decays (see e.g.  \cite{BUFLE1998,BFRS2004}). In this case the  typical energy of decaying $B$ mesons is of the order $m_b$, so the good knowledge of the  infrared value of the top quark mass is important for the description of heavy meson decays. Of course, the knowledge of mass at high scales can useful for model builders. However, in the nonperturbative treatment here we are mainly for the physic not far above the electroweak scale, the knowledge of the quark mass at higher scales can be useful for model builders as well.

The last but not least motivation  is a direct check the effect of higher order corrections including Higgs trilinear coupling on the solution. To do this the appropriate  two loop skeleton diagram is calculated and included into the top quark SDE. These, and other details of the model are  described in the Section II. and Section III. of presented paper.    
To find the correct solution in the full Minkowski space is a problematic task for a strong coupling theory like QCD.  First we solve quark SDE in Euclidean space by  standard numerical manner in the Section IV.  In Section V. we continue the solution to the timelike axis in a way that experimentally known pole mass is achieved by the correct solution. This is achieved by  resolving of the SDE with   renormalized mass adjusted to obtain the correct physical pole mass at the end.

\section{Schwinger-Dyson equation for top quark mass function}

Neglecting the weak interaction, the quark propagator $S$ can be conventionally  characterized by two independent scalars, the mass function $M$ and the renormalization wave function $Z$ such that
\be
S(p)=\frac{Z(p)}{\not p-M(p)}\, .
\ee

The SDE for the inverse of $S$ reads
\bea \label{zdar}
S(p)^{-1}&=&\not p -g_Y<\phi>-\Sigma_A(p)-\Sigma_h(p)+...
 \\
\Sigma_A(p)&=&i g^2\int\frac{d^4q}{(2\pi)^4}\Gamma_{\alpha}(q,p)
G^{\alpha\beta}(p-q)S(q)\gamma_{\beta} \,
\nn \\
\Sigma_h(p)&=&i g_Y^2\int\frac{d^4q}{(2\pi)^4}\Gamma_h(q,p)
G_h(p-q)S(q) \, \nn
\eea
where $g_Y$ is the top Yukawa coupling, Higgs vev  $<\phi>=246\, GeV/\sqrt{2}$ and $g$ is QCD gauge coupling. The dots represent omitted contributions, e.g. $W,Z,\gamma$ and related Goldstone exchanges.
 $G,\Gamma$ stand for boson propagators and the top quark-boson vertices and they satisfy their own SDEs. 
\begin{figure}
\centerline{\epsfig{figure=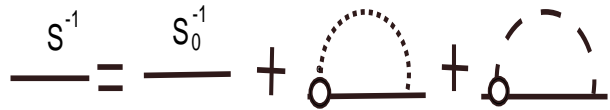,width=8truecm,height=1.5truecm,angle=0}}
\caption[caption]{ Diagrammatic representation of top quark SDE. The solid line stands for quark and the dot (dashed) line stands for the gluon (Higgs) respectively. The circles represent full vertices.  } 
\end{figure}

 The knowledge about these Greens function   is necessarily  limited due to   theoretical and experimental reasons. They need to be approximated if they are not selfconsistently contained in a given   truncation scheme of the SDEs system. A natural treatment of this problem is to make an expansion  in the number of loops. Performing such an expansions for vertices $\Gamma=\sum_i \Gamma^{i}$ and substituting this into the selfenergy (\ref{zdar}), one gets the expansion for the mass function. Explicitly, the loop expansion for the selfenergy in (\ref{zdar}) should read
\be
\Sigma_{A}=\Sigma_A^{[1]}+\sum_i \Sigma_A^{[i+1]}\,
\ee
and similarly for  $\Sigma_{h}$.
% note this is independent on the vertex Ansatz. 

In the simplest approximation the 
first order estimates can be obtained by using the classical vertices 
\bea \label{skeleton}
\Sigma_A^1(p)&=&i g^2\int\frac{d^4q}{(2\pi)^4}\gamma_{\alpha}
G^{\alpha\beta}(p-q)S(q)\gamma_{\beta} \,
\nn \\
\Sigma_h^1(p)&=&i g_Y^2\int\frac{d^4q}{(2\pi)^4}
G_h(p-q)S(q) \, 
\eea
where the all propagator functions entering the Eqs. (\ref{skeleton}) are fully dressed.

Including "radiative corrections" to the SM model Higgs
one should get coupled SDEs for $G$ and $S$.  In the case of light Higgs, the top-antitop  quark loop contribution would lead to  the extremely large negative contribution to the Higgs boson mass. This  mass hierarchy problem, although formally solved by renormalization, is one of the main motivation for extension of the Standard Model and the reason why the SM is regarded as an effective low energy theory. In the extensions of SM the mass hierarchy is stabilized by the introduction of the other scalars  \cite{SING1,SING2,SING3}, SM doublets \cite{DOUB1,DOUB2}, or  is  eliminated by supersymmetry or the Lee-Wick  SM modification \cite{GROCWI}. In all these models,  a new particle content is expected at few TeV, the quadratic divergences to Higgs mass  are reduced and the free propagator could be a reasonable approximation of the exact Higgs propagator for a broad regime of scales.  Therefore the simplest -free  Higgs boson propagator: 
\be \label{barek}
G_h(p)=\frac{1}{p^2-m^2}\, 
\ee
is used, where $m$ is the physical Higgs boson mass (\ref{higsmass}).

Following the recent precision test of the Standard Model \cite{LEP}.
the analysis of the radiative corrections favor a light Higgs boson $m\simeq 76 \, GeV $. 
Because of the  lack of an experimentally observed Higgs particle,
 the mass of the Higgs boson could be rather close  to the experimental lower bound  $m>114.4 \,GeV$ \cite{BARATE}. In this paper the following value of the Higgs mass is chosen
\be \label{higsmass}
m=120 \,GeV,
\ee 
as the input parameter in our model.

 At low scales,  $q \simeq \Lambda_{QCD}$, the running QCD coupling is large and the  dressing of the gluon-quark-antiquark vertex can play an important role in the description of  light flavor dynamics \cite{AFES2008}.  However, in the case of the top quark, the running coupling becomes quite small $\alpha_{QCD}(M)\simeq 0.1$ and one can economically  include the contribution of higher orders to the  effective running coupling. For this purpose  the following prescription for the SDE QCD-part kernel is used:
\be  \label{aprc}
g^2G^{\mu\nu}(k)\Gamma_{\nu}(q,p)\rightarrow 
4\pi\alpha(k^2,\Lambda)\frac{-g^{\mu\nu}+\frac{k^{\mu}k^{\nu}}{k^2}}{k^2+\ep}
\gamma_{\nu}  \,. 
\ee

where  $\alpha$ represents the analytical running coupling  \cite{SOL1,SOL2,SOL3,SOL4}.
 In the one loop approximation  it is given by the following expression: 
\be  \label{rep}
\alpha(q^2,\Lambda_{QCD})=\int_0^{\infty} d\omega\, \frac{\rho_g(\omega,\Lambda_{QCD})}{q^2-\omega}\, ,
\ee 
where 
\be \label{run}
\rho_g(\omega,\Lambda_{QCD})=\frac{4\pi/\beta}{\pi^2-\ln^2{(\omega/\Lambda^2_{QCD})}} \, .
\ee
Recall that the analytical running coupling is  constructed in  a simple way that  avoids the unwanted artifact of perturbation theory- the Landau pole at $q^2=\Lambda^2_{QCD}$- which is subtracted away and thus the running coupling is free of unphysical singularities. The procedure has been  generalized to higher orders, provided that the ultraviolet asymptotic behaviour of such running constant is identical with the  perturbative  result.  In the numeric here the  one loop
approximation (\ref{run}) is used  with the numerical value of $\Lambda_{QCD} $, $ \Lambda_{QCD}= 500MeV$ for six active quarks.  The beta function coefficient  is thus
\be
\beta=\frac{11N_c-2N_f}{3}
\ee
with $N_c=3, N_f=6$.

The computation is carried out in Landau gauge and  the $Z=1$ approximation is used. Whilst in pure gauge theory the effect of the $Z=1$ approximation can be minimized by proper adjustment of the gauge fixing, the importance of $Z$  in the presence of the Higgs field is not explored and  remains to be estimated in a future study.

%At this point let as emphasized that the recent solutions of Landau gauge gluon propagator from SDEs (e.g. in background field gauge) approximately agree with lattice results. Up to a certain difference,  the resulting gluon propagator starts to copy its known perturbation behaviour at few GeV. We simulate this deeply infrared effect and  have also studied the possible influence by explicit introduction of small gluon mass $M\simeq 0.5-1GeV $ with  simultaneous enhancement of gluon form factor at the scale of few GeV ( by enhancing $\rho$ appropriately). We have obtained negligible changes for the running top quark mass in this case. It exhibits sufficiency of our running QCD  approximation given by \ref{run} 
%in the all range of momentum scales, and such infrared nonperturbative changes in  the SDE kernel are not further considered. 

\section{Solving top quark SDE in Euclidean space}

Using the following formula 
\bea \label{hangar}
&&\int_0^{\pi} \frac{d\eta}{\pi} \frac{sin^2\eta}{l^2-2|l||p|cos\eta+p^2+m^2}=
\nn \\
&&\frac{-p^2-q^2-m^2+\sqrt{(p^2+q^2+m^2)^2-4p^2 q^2}}{-4p^2} \, ,
\eea
the  angular integrations in one loop skeleton diagram in (\ref{skeleton}) can be easily evaluated.
After the explicit integration the Higgs-top loop contribution  can be cast into the one dimensional integral  
\be \label{h1}
\Sigma_h^{[1]}(x)=\frac{\alpha_Y}{\pi}
\int_0^{\infty}d y \, \frac{M(y)}{y+M^2(y)}
K(x,y,m^2)\,\, ,
\ee
where $\alpha_Y=g^2_Y/4\pi$ and the functions $K$ is defined as 
\be
K(x,y,z)=\frac{2y}{x+y+z+\sqrt{(x+y+z)^2-4x y}} \label{reg2}\,.
\ee

Likewise, for the one loop QCD contribution we get 
\be  \label{A1}
\Sigma_A^{[1]}(x)=\frac{1}{\pi}
\int_0^{\infty}d y \, \frac{M(y)}{y+M^2(y)}
V(x,y)\,\, ,
\ee
with the function $V$ defined as 
\be
V(x,y)=-\int_0^\infty d \omega \frac{\rho_g(\omega)}{\omega}
 \left[ K(x,y,0)-K(x,y,\omega)\right]\, .
\ee

\begin{figure} 
\centerline{\epsfig{figure=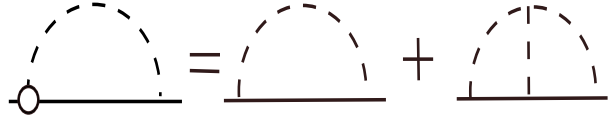,width=8truecm,height=1.5truecm,angle=0}}
\caption[caption]{ Diagrammatic representation of selfenergy contribution due to the Higgs. The circle stands for the full vertex, the rhs. is the approximation employed here.  } \label{woena}
\end{figure}

In addition, the one loop skeleton contribution to the 
Higgs-quark-antiquark proper vertex  (see Fig. \ref{woena}) 
is carefully included. This is equivalent to the two loop 1PI contribution for the  top quark dynamical mass function which reads:
\be  \label{h2}
\Sigma_h^{[2]}(p^2)=\lambda v g_Y^3 I(p) \, ,
\ee
where $\lambda/4$ is the Standard Model quartic Higgs coupling  
and $I(p)$ is the following two loop integral:
\bea \label{zdarec}
&&I(p)=\frac{Tr}{4}
i \int\frac{d^4l}{(2\pi)^4}\frac{1}{(l-p)^2-m^2}
\frac{ \not l+M(l)}{l^2-M(l)^2}
\nn \\
&\times& i \int\frac{d^4q}{(2\pi)^4}\frac{1}{(q-p)^2-m^2}
\frac{ \not q+M(q)}{q^2-M(q)^2}
\nn \\
&\times&\frac{1}{(q-l)^2-m^2} \, .
\eea
After the Wick rotation to  Euclidean space, the integrations in (\ref{zdarec}) are not calculable directly, however most of them  can be  calculated analytically by performing just one quite standard angular approximation  (\ref{zdarec})). This approximation eliminates the angle between the two internal loop momenta in the following manner:
\be
(l-q)^2\rightarrow l^2\theta(l^2-q^2)+q^2\theta(q^2-l^2)
\ee
and so writing also for $l.q$ product (this stem from Dirac trace)  
\bea
l\cdot q&=&\frac{l^2+q^2-(l-q)^2}{2}
\nn \\
&\rightarrow& \frac{q^2}{2}\theta(l^2-q^2) +\frac{l^2}{2}\theta(q^2-l^2)\, ,
\eea
the expression for  $I$ can be recast as:
\bea \label{zdar1}
I(p)&=& \int\frac{d^4l}{(2\pi)^4}\frac{1}{(l-p)^2+m^2}
\int\frac{d^4q}{(2\pi)^4}\frac{1}{(q-p)^2+m^2}
\nn \\
&\times&\frac{\frac{M(l)M(q)+q^2/2}{l^2+m^2}\theta(l^2-q^2)+\frac{M(l)M(q)+l^2/2}{q^2+m^2}\theta(q^2-l^2)}{(q^2+M^2(q))(l^2+M^2(l))}\, \nn.
\eea
Here, it is an opportune point to remark that such an angular approximation has been  extensively used in phenomenological SDE studies of QCD and QED4 even at one loop level. In our case the coupling constant is small enough and following the  critical one loop analysis performed in \cite{RC1990}, this must be a reliable approximation in our two loop case. In Euclidean domain it can lead to a few percent error in $I$. As we have estimated posterior, it  entails only a tiny (a few promile) error in the total result for $M$.

Using the  formula (\ref{hangar}) the remaining angular integrations can be performed, resulting  the following expression for $I$:
\bea \label{zdar2}
&&I(p)= \int\frac{dq^2}{8\pi^2}\frac{p^2+q^2+m^2-\sqrt{(p^2+q^2+m^2)^2-4p^2 q^2}}{p^2[q^2+M^2(q)]}
\nn \\
&&\times \int\frac{dl^2}{8\pi^2}\frac{p^2+l^2+m^2-\sqrt{(p^2+l^2+m^2)^2-4p^2 l^2}}{p^2[l^2+M^2(l)]} 
\nn \\
&&\times\frac{\frac{M(l)M(q)+q^2/2}{l^2+m^2}\theta(l^2-q^2)+\frac{M(l)M(q)+l^2/2}{q^2+m^2}\theta(q^2-l^2)}{(q^2+M^2(q))(l^2+M(l)^2)} \nn \, .
\\
\eea
In what follows we  interchange of the variables $l\leftrightarrow q$ in the second term of the third line of the Eq. (\ref{zdar2}). Considering the  appropriate  prefactors, $I(p)$  can be finally written in the following way:
\bea \label{zdar3}
I(p)=\int_0^{\infty}dq^2\frac{p^2+q^2+m^2-\sqrt{(p^2+q^2+m^2)^2-4p^2 q^2}}{p^2[q^2+M^2(q)]}&&
\nn \\
\times \int_0^{q^2}dl^2\frac{p^2+l^2+m^2-\sqrt{(p^2+l^2+m^2)^2-4p^2 l^2}}{p^2[l^2+M^2(l)]}&&
\nn \\
\nn \times\frac{1}{32\pi^2}\frac{l^2/2+M(l)M(q)}{(q^2+m^2)(q^2+M^2(q))(l^2+M^2(l))}&&
\\
\eea
\begin{figure}
\centerline{\epsfig{figure=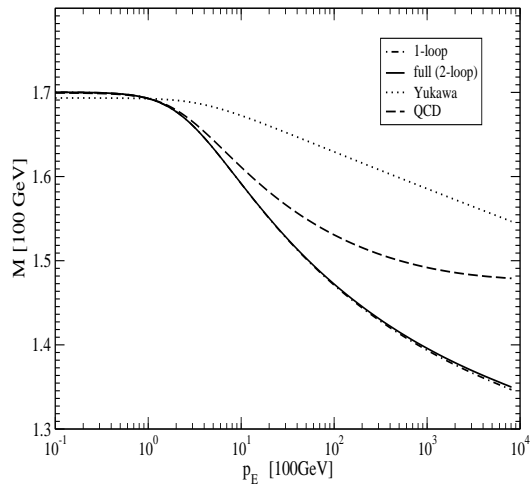,width=8truecm,height=8truecm,angle=270}}
\caption[caption]{ Running  Top quark mass as described in the text. The solid line represents the full solution, dot-dot-dashed line stands for the case when two loop skeleton is omitted. The dashed and dotted lines stand for pure QCD and Yukawa solutions respectively. } \label{details}
\end{figure}

Putting these all together, the SDE for top quark mass function that is to be solved reads
\be \label{zaver}
M(x)=g_Y<\phi>+\Sigma_A^{[1]}(x)+\Sigma_h^{[1]}(x)+\Sigma_h^{[2]}(x), 
\ee
where the individual terms are given by (\ref{A1}), (\ref{h1}) and (\ref{h2}) wherein  $I$ is given by Rel. (\ref{zdar3}). As a consequence of the $Z=1$ approximation the function  $M(x)$ is renormgroup invariant.  After making a subtraction the "renormalized" equation  actually solved reads
\bea \label{uzaver}
M(x)&=&M(\xi)+\Sigma(x)-\Sigma(\xi)
\nn \\
\Sigma(x)&=&\Sigma_A^{[1]}(x)+\Sigma_h^{[1]}(x)+\Sigma_h^{[2]}(x) \, ,
\eea
where the renormalized mass at the scale $\xi$ is related to the bare top quark mass through the following rel.: $M(\xi)=g_Y<\phi>+\Sigma(\xi)$.

\section{Results in spacelike regime}

In this section, we discuss the numerical solution of the renormalized Euclidean SDE (\ref{uzaver}).
The physical  mass pole being on the timelike axis cannot be directly used for the solution.
The main purpose of this section is to exhibit the importance of various contribution in the case 
of light Higgs exchanges.

The  SDE  (\ref{uzaver}) has been solved  by the method of  iterations with high accuracy.
For this purpose  we have chosen the (spacelike) renormalization scale to be
\be
\xi^2=100 GeV^2 \,
\ee
and  fixed the renormalized mass $M(\xi)$ through the Yukawa coupling.

The two loop diagram depicted in Fig. \ref{woena}  includes the triplet Higgs interaction constant, 
which is already determined through the quartic one. In our numerical calculation the coupling constant actually used is read from the relation $\lambda=\frac{m^2}{2v^2} $ (at the given scale $\xi$).

The resulting mass function is displayed in the Fig. \ref{details}.  The presented calculations were performed with $g_Y=0.977$, so the
corresponding renormalized mass is adjusted so that  $M(100GeV)=169.25GeV$. With these inputs we get the numerical solution. As the presented solution is regularization independent, we used hard cutoff  regulator $\Lambda>>M(0)$ obtaining  the same solution  when $\Lambda$  was varied through   many orders. The mass function is increasing when going to infrared, reaching its infrared value 
 $M(0)=170.00GeV$, being not  far the experimental one. How to gain the solution actually based on determined physical top quark mass will be discussed in the section. Before this we discus some general features of the solution.

In our presented framework of SDEs the dynamical mass function is  slowly varying function in the infrared. Up to few GeV contribution the infrared mass does not change drastically at the scale of 0-100 GeV.

The Yukawa interaction between Higgs and top quark is quite strong even when comparing to the QCD interaction strength. In Fig. \ref{details}. we show the comparison of solutions stemming purely from the  Yukawa interaction and from the pure QCD  (by switching off  QCD  or Yukawa interaction). The same value of the renormalized top quark mass is kept for this purpose. As expected, the QCD  dominates in the infrared regime, while in high momenta, $q^2>>M_t^2$ both interactions are of the same magnitude.

Interestingly, the two loop effect with Higgs trilinear coupling gives a marginal  contribution  for all $p^2$. Numerically, two loop skeleton effect is  comparable with the one loop PT electroweak corrections. For a heavier Higgs the one loop Higgs contribution becomes less important, while the two loop contribution appears to be less affected since the triplet Higgs coupling is getting strong. 
We have also solved the SDE with different Higgs masses as well. 
For instance Higgs heavy as $m= 0.5TeV$, two loop contribution  becomes more important giving rise  a few $GeV$ negative contribution in the infrared top quark mass. However, one should  note that in  this  case the Higgs sector becomes strongly interacting what would require more careful reinvestigation due to the new  nonperturbative dynamics \cite{RUP1,RUP2,RUP3}.

\section{Solution for all momenta}

Experimentally the top quark mass is reconstructed by collecting jets and leptons. From the position of the bunch in cross section measured at the Tevatron  the pole position is identified $M_t=172.4\pm1.4$. The ambiguity and uncertainty of the full top propagator pole mass is affected by experimental methods and theoretical weaknesses of perturbation theory description of  jets, e.g. reconciling the contribution of soft and collinear particles. Furthermore, the correct identification of the mass requires nonperturbative technique at all. While, including perturbative 1-loop $b,W$ electroweak correction this pole could only move into the second sheet complex plane giving rise top quark decay width $\Gamma_t= 1.5 GeV$, the perturbation theory cannot give nonambiguous result because of uncertainty proportional to $\Lambda_{QCD}$ \cite{lochneskaodr2,BIUR94,BEBR94}. The real pole of the (pure QCD) perturbation theory can turn to be complex because of confinement phenomena  as recently  observed  in   \cite{SAU8}  by studying complex mass generation in temporal Euclidean space.    
  
In this paper we do not solve the problem of confinement of th top quark in SDE framework, instead  we show that the running top quark mass function turns to be stable, very slightly varying, quantity when continued to the timelike momenta. For this purpose the mass function $M(-x)$ at timelike square of the fourmomenta $t=-x>0$ is constructed by the continuation of the left hand side of SDE by taking $x\rightarrow -t$.
We can write for the continued solution

\bea \label{timelike}
M(t)&=&M(\xi)+\Sigma(t)-\Sigma(\xi)
 \\
\Sigma(t)&=&\Sigma_A^{[1]}(t)+\Sigma_h^{[1]}(t)+\Sigma_h^{[2]}(t) \, ,
\nn \\
\Sigma_h^{[1]}(t)&=&\frac{\alpha_Y}{\pi}
\int_0^{\infty}d y \, \frac{M(y)}{y+M^2(y)}
K(-t,y,m^2)\,\, ,
\label{rhsofmass} \\
K(-t,y,z)&=&\frac{2y}{-t+y+z+\sqrt{(-t+y+z)^2+4t y}} \, ,
\nn
\eea
and similarly, the functions $\Sigma_A^{[1]},\Sigma_h^{[2]}$ are obtained by the substitution 
$x\rightarrow -t$ in their kernels.

Since the mass function on the rhs. of Eq. (\ref{rhsofmass}) remains defined at the spacelike regime, the pole mass cannot be used as the renormalized point directly. To achieve the solution of SDE with experimentally known value of top quark mass, we shift the renormalized mass in  (\ref{uzaver}) and then have a look for the solution for $M_t$  by integrating the equation (\ref{timelike}). With sufficient accuracy it is easily achieved by hand  iteration process.

\begin{figure}
\centerline{\epsfig{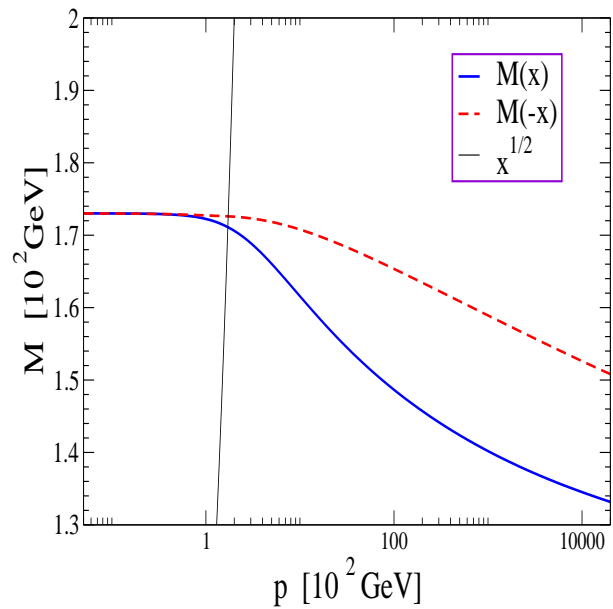}}
\caption[caption]{ Running  top quark mass based on observed Tevatron pole mass. The solid line represents the spacelike and dashed the timelike solution, dotted solid line represents the linear function $f=\sqrt(x)$ ,which when cuts the dashed line, identifies the real pole of the full top quark propagator).  } \label{final}
\end{figure}

The experimentally observed  mass  knowledge based solution is presented in Fig.
\ref{final}. The numerical value $M_t=172.6GeV$ is obtained as the solution for pole mass. The timelike solution is plotted at the negative axis. The resulting Yukawa coupling to our $120GeV$ heavy Higgs field has been adjusted as $g_Y=0.9845$. The solution is real everywhere as the method is inefficient to provide absorptive part from the nonanalytical cut at real axis at $M_t$. The experimental uncertainty  defines the errors repesented by  narrow band of width $\sim 2-2.8 GeV$  with  presented solution inside. We do not display these.

 The other interesting values we can quote here are (in GeV):
\be
M(10^{12})=134.5; M(M_t^2)=171.0; M(M_{Z}^2)=172.4; M(0)=173.0; M(-M_t^2)=172.6,  M(-10^{12})=151.2,  \nn \, .
\ee

\section{Conclusion}

The SDE calculation of running top quark mass previously discussed in the literature \cite{SMJAKA1994} is presented in some extent. The obtained solution is based on the measured physical top quark mass. 
The top quark mass can be safely evolved to small $q^2$ when one avoids the pathology of perturbation theory, e.g. Landau pole in gluon propagator.
It exhibit great stability at all scales of spacelike and timelike domain as well. For the timelike domain the function is such slowly varied that the top quark physical mass appears to be a rather good approximation at all low scales.

  At low scales, QCD contribution  dominates over the one due to the Higgs loop(s), at large $q^2$
both Higgs and QCD loops are comparable. 
In addition, we  estimated the  two loop Higgs contribution,  which gives only tiny contribution for the case of the light Higgs. The extension of presented technique to the more general models, e.g. with more Higgs doublets and/or scalar singlets added to SM Higgs sector is straightforward.

\onecolumngrid

%%%%%%%%%%%%%%%%%%%%%%%%%%%%%%%%%%%%%%%%%%%%%%%%%%%%%%%%%%%%%%%%%%%%%%%

\end{document}